\documentclass[letterpaper,10pt]{article}

\usepackage[square]{natbib}
\usepackage{graphicx}   
\usepackage{epsfig}
\usepackage{amssymb}

\begin{document}

\newcommand{\avk}{\langle k \rangle}
\newcommand{\fluck}{\langle k^2 \rangle}

\title{{\bf Random Walks on Directed Networks: the Case of PageRank}}

\author{Santo Fortunato$^{1,2,3}$, A. Flammini$^{1}$}

\maketitle 

\begin{center}
\small{$^1$ School of Informatics and Biocomplexity Institute,
Indiana University, Bloomington, IN, USA}\\
\small{$^2$ Complex Networks
Lagrange Laboratory (CNLL), Institute for Scientific Interchange
(ISI), Turin, Italy}\\
\small{$^3$ Fakult\"at f\"ur Physik, Universit\"at Bielefeld, D-33501 Bielefeld, Germany }
\end{center}

\begin{abstract}

PageRank, the prestige measure for Web pages used by Google, 
is the stationary probability of a peculiar random walk on directed graphs,
which interpolates between a pure random walk and a process where 
all nodes have the same probability of being visited. 
We give some exact results
on the distribution of PageRank 
in the cases in which the damping factor $q$ approaches the two limit values
$0$ and $1$.
When $q\rightarrow 0$ and for several classes 
of graphs the distribution is a power law 
with exponent $2$, regardless of the in-degree distribution. When $q\rightarrow 1$ 
it can always be derived from
the in-degree distribution of the underlying graph, if the out-degree is the same
for all nodes.

\end{abstract}

%%%%%%%%%%%%%%%%%%%%%%%%%%%%%%%%%%%%%%%%%%%%%%%%%%%%%%%%%%%%%
\section{Introduction}

Since the letter of Pearson~\citep{pearson}, published on Nature in 1905, 
random walk has become a central concept in many branches of the physical sciences.
The number of applications and studies dedicated to the subject is in fact so large
that to give even a very partial list of references is an overwhelming enterprise
(see~\citep{hughes} for a recent and fairly complete review).
Most of the attention, so far, has been devoted to the study 
of random walks and related stochastic processes on $d$-dimensional
euclidean spaces and regular lattices, for their obvious relevance to
physical problems. To extend the definition of random walk 
to an arbitrary graph is trivial, but its study is relatively less 
developed. In this paper we address the issue of the stationary 
probability of a random walk on a directed scale-free graph. 

The specific 
application we have in mind is the study of Pagerank (PR), the prestige measure
that the search engine Google (and several other search engines)
employs to measure  
the prestige of Web pages. When a user submits a query, 
the hits returned by Google are 
ranked according to their PR values. 
As it will be clear in a moment, such a measure
is the stationary probability of a random walk  
on the Web graph, where each node represents a Web page and edges represent
the hyperlinks (naturally directed) connecting the pages. 

\begin{figure}[t]
\begin{center}
    \includegraphics[width=5.5cm]{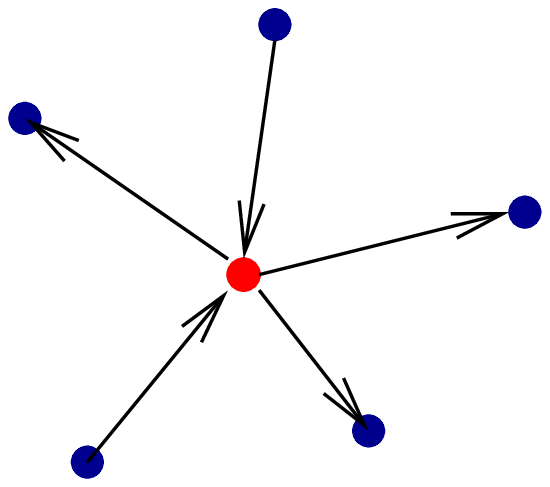}
\end{center}
    \caption{\label{fig1} The node in the center has two incoming and three outgoing links.}
\end{figure}

Let us consider an arbitrary undirected graph and a random
walker moving on it. At any (discrete) time step the walker jumps from
the node where it is sitting on to one of its neighbors chosen with equal 
probability. It is trivial to show that, at stationarity, the probability
of each node to be visited is proportional to its degree, i.e. the number
of neighbors of the node.
If the graph is directed we have to distinguish (see Fig.~(\ref{fig1})) 
the links adjacent to a node in
incoming (those that point to the node) and outgoing (those that point away
from it).
If the random walker is allowed to follow 
only the outgoing links from the node where it presently is, 
the problem of finding the stationary probability is far more complicated.
Such probability will in general depend on the overall topological organization 
of the graph itself, and cannot be expressed in terms of simple topological
quantities like the degree of a node.
In fact, due to the directedness of the links, the graph may have regions
that the walker can enter in but not escape from. The stationary probability
will be trivially concentrated in these regions. In order to prevent this 
from happening we will consider a modified (directed) random walker whose behavior 
is defined by the following two rules: 

\begin{itemize}
\item{with probability $1-q$ the walker follows any outgoing link of $i$, chosen 
with equal probability;}
\item{with probability $q$ it moves to a generic node of the network (including $i$), 
chosen with equal probability.}
\end{itemize}

This will suffice to ensure a non-zero stationary probability on every
node. When considered in the context of the Web graph, the process described
above could be thought as a rough modelization of a Web surfer that 
occasionally (with probability $q$) decides to interrupt his/her browsing
and to restart it from a randomly chosen page.
The stationary probability of this process is 
exactly PR.
To adhere to the computer science terminology, we will
refer to the probability $q$ as to the damping factor. The damping factor
adopted in real applications is generally small ($q \sim 0.15$).
  
\begin{figure}[t]
\begin{center}
    \includegraphics[width=9cm]{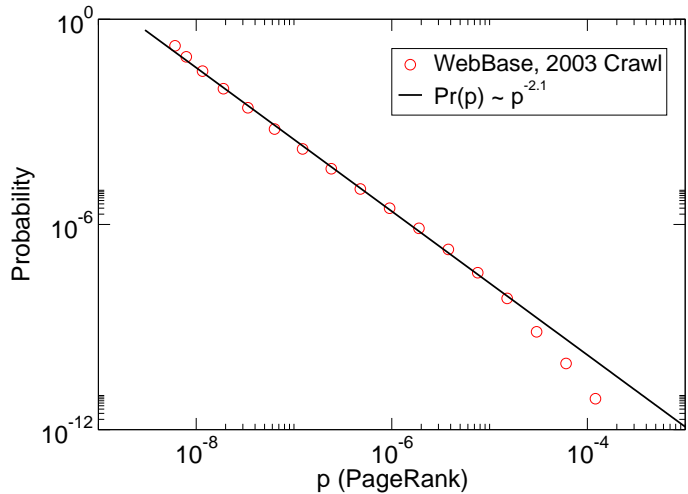}
\end{center}
    \caption{\label{figPR} PR distribution for a large sample of the Web graph,
produced by the WebBase collaboration in 2003
(\texttt{\small{www-diglib.stanford.edu/$\sim$testbed/doc2/WebBase/}}). 
The damping factor is $q=0.15$.} 
 \end{figure}

\begin{figure}[t]
\begin{center}
    \includegraphics[width=9cm]{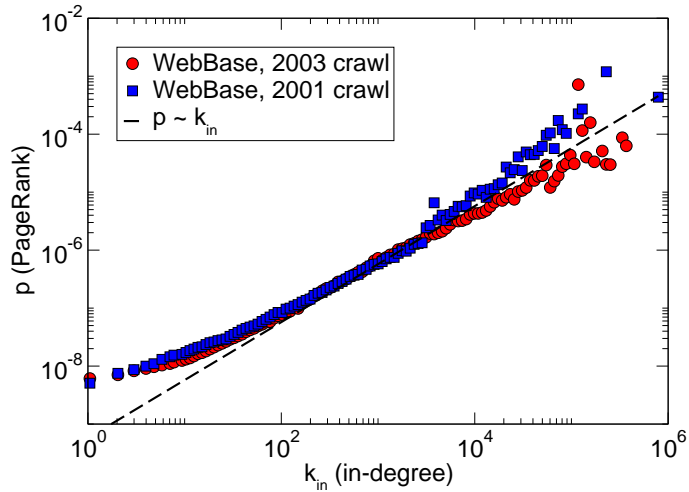}
\end{center}
    \caption{\label{aaa} PR versus in-degree for two samples of the Web graph,
produced by the WebBase collaboration in 2001 and 2003. 
The damping factor is $q=0.15$.} 
 \end{figure}

A brave analogy with the undirected case could lead to the hypothesis that PR
is roughly proportional to the in-degree of a node (number of incoming links), modulo
corrections due to the small damping factor. Such a view could be 
further supported by the observation that the distribution of PR
for the real Web has a power law decay~\citep{Upfal2002} characterized by an exponent
 $2.1$ (see Fig.~(\ref{figPR})), like the distribution of the in-degree~\citep{Baradiam}
(note that, when referring to the Web and unless otherwise specified, we always assume
a damping factor of $q \sim 0.15$). 
A direct measure of PR versus in-degree on two large samples
of the Web graph is shown in Fig.~(\ref{aaa}), where the value of PR has been averaged
over nodes with the same in-degree. The plot exhibits an almost linear behavior
with deviations at small degrees, when the effect of the damping factor is more relevant.
Mean field calculations show that there is a positive correlation between
PR and in-degree~\citep{santoMF} and a linear relation between in-degree and the mean PR
for nodes of equal in-degree can be safely assumed if the degree correlations between
adjacent nodes are weak.  
On a generic directed graph, the linear relationship between PR (even if considered on average) 
and the in-degree is not granted and it depends on the global organization of the
graph itself. To address the issue of PR distribution for an arbitrary graph and a generic
$q$ would therefore require a case by case study. In this paper, therefore, we concentrate on
the two interesting limits, i.e. $q \rightarrow 0$ and $q \rightarrow 1$, that show some degree
 of universality.
In these two limits it is possible
to derive analytical expressions for the distribution of PR. For small $q$-values,
a master equation approach allows us to solve the problem for 
special classes of networks. For $q\rightarrow 1$, it is possible to establish a one-to-one
correspondence between the distribution of PR and that of in-degree, as long as
the number of outgoing links from each node (out-degree) is the same.
Further, to have a better control on the topological characteristics of the graph
and how they correlate with the PR distribution we work 
with graphs generated by random processes or processes of growth. 

\section{PageRank}

Let us consider a generic directed network with $n$ nodes.
Let $p(i)$ be the PR of node $i$.
The vector $p$ satisfies the following self-consistent system of relations:
\begin{equation}
p(i)=\frac{q}{n}+(1-q)\sum_{j: j \rightarrow i}\frac{p(j)}{k_{out}(j)} \hskip0.9cm i=1,2,
\dots, n
\label{eq1}
\end{equation}
where $j \rightarrow i$ indicates a
link from $j$ to $i$ and $k_{out}(j)$ is the out-degree of node $j$.
In the following we always assume that each node has at least one outgoing link, and therefore
Eq.~(\ref{eq1}) is well defined.
To compute $p$ amounts to solve the eigenvalue problem for the 
transition matrix $\cal M$, whose element
${\cal M}_{ij}$ is given by the  expression:
\begin{equation}
{\cal M}_{ij}=\frac{q}{n}+(1-q)\frac{1}{k_{out}(j)}A_{ji},
\label{eq2}
\end{equation}
and where $A$ is the adjacency matrix of the graph ($A_{ji}=1$ if there is a link from $j$ to $i$,
otherwise $A_{ji}=0$). 

The stationary probability of the process described by $\cal M$ is given by its 
principal eigenvector. Its calculation is a standard problem of numerical analysis and 
can be achieved by repeatedly applying the matrix $\cal M$ to a generic vector $p_0$ not
orthogonal to $p$. It is easy to show, in fact, 
that $1=\lambda_0>\lambda_1\geq ...\geq\lambda_n$ 
($\lambda$'s being the eigenvalues of $\cal M$), and therefore 
$\lim_{l \rightarrow \infty}{\cal M}^l p_0 = p$. The powers of the matrix
$\cal M$ introduce powers of the eigenvalues $\lambda$ in the decomposition of $p_0$ 
in eigenvectors of $\cal M$. In this way, if we aim at calculating $p$ with an accuracy 
$\epsilon$, the vector ${\cal M}^lp_0$ delivers $p$ with corrections 
at most of the order $\lambda_1^l$,
so we can safely stop the procedure when $\lambda_1^l\sim\epsilon$, i.e. when 
$l\sim log(\epsilon)/log(\lambda_1)$.
In this way, the method converges rather quickly:
in practical applications, it turns out that less than one hundred iterations suffice to  
calculate the PR of a network with $10^7-10^8$ vertices.

PR, and therefore its distribution, depends on the damping factor $q$ in a non-trivial way.
A first rigorous investigation of this problem was presented in \citep{boldi}
with focus on how the ranking of pages is influenced by changing $q$ and where 
some close expressions for derivatives of PR with respect to $q$ were derived. 
The damping factor can be considered as an interpolation parameter between a simple
random walk and a pure scattering process. When $q=0$, the process reduces to a simple 
random walk, and one may end up with a trivial invariant measure concentrated
on a small subset of nodes. When $q=1$, the walker can jump to any node at each step, 
with probability $1/n$. The PR of
all nodes is then the same, and equals $1/n$, as one can see by setting $q=1$ in Eq.~(\ref{eq1}).
The distribution of PR is therefore a Dirac $\delta$ function centered at $1/n$.
For $0<q<1$ the distribution is not trivial, and in general it strongly depends on 
the underlying graph. 
On the other hand, in the two limits $q\rightarrow 0$ and $q\rightarrow 1$, Eq.~(\ref{eq1})
assumes forms which lend themselves to simple analytical derivations and the PR 
distribution can be
exactly determined for a large set of graphs. 
It is worth remarking that the limit of small $q$
is the relevant one for Web applications.

\section{The general case of a direct-loopless graph}

Given a generic directed graph, the PR of a specific node depends on the overall
arrangement of the graph and cannot be calculated on the basis of local properties only. 
In the following we focus on a wide, although more restricted class of networks
for which analytical solutions are possible. To this class belong networks obtained
through a growth process that are particularly important for real world applications.
 
Let us label the nodes of the network $1$, $2$, ..., $n$. We assume that if an oriented 
path from node $i$ to node $j$ exists, there is no path from $j$ to $i$ (in other words,
it is impossible to get back to a given starting point following an oriented path of the graph).
Networks that result from a growth process, where new nodes are introduced at discrete 
time steps together with their new oriented links, obviously belong to this class.
In fact, we can label nodes according to their age (node 1 being the oldest) so that
a directed link between $i$ and $j$ may exist only if $i>j$.

It is easy to verify that the PR $p(i)$ of a generic node $i$ of a graph in this class
can be written as follows:
\begin{equation}
\label{equA}
p(i)= \frac{q}{n} \Big[ 1 + \sum_{j =1,n} \sum_{l^{ij} \in L^{ij} } 
\frac{ (1-q)^{d(l^{ij})} }{\prod_{s=1,d(l^{ij})} k(l^{ij}_s) } \Big]
\end{equation} 
where the first sum runs over all nodes in the graph and the second over all paths from
a generic node $j$ to node $i$. Each path in the second sum is weighted by as many factors 
$(1-q)$ as links along the path ($d(l^{ij})$ is the lenght of path $l^{ij}$)
and is also weighted by the inverse of the degree ($k(l^{ij}_s)$) of each node $s$ encountered 
along the path.
Although correct for any $q$, the formula above is not very transparent. In order
to get some understanding on the expected distribution of Pagerank in a graph,
we specialize Eq.~(\ref{equA}) to the case in which each node has a fixed number $m$ of 
outgoing links. We further focus on the limit $q \rightarrow 0$ that has an immediate
interpretation in terms of walks over the graph. Under the assumptions above,
the expression for the PR of a node $i$ simplifies to

\begin{equation}
\label{equB}
p(i)= \frac{q}{n} \Big[ 1 + \sum_{j =1,n} \sum_{l^{ij} \in L^{ij} } 
\Big( \frac{ 1-q}{m}\Big)^{d(l^{ij})}\Big]
\end{equation} 

In the following we show that if the graph is grown according to preferential 
attachment or copying mechanism and $q$ is sufficiently small, we should expect 
an algebraic distribution of PR characterized by an exponent $2$,  
for any value of $m$. We will give the proof in the case $m=1$ and a hint to a general proof
in Sec.~\ref{hint}.

\subsection{The limit $q \rightarrow 0$}

\begin{figure}[t]
\begin{center}
    \includegraphics[width=5cm]{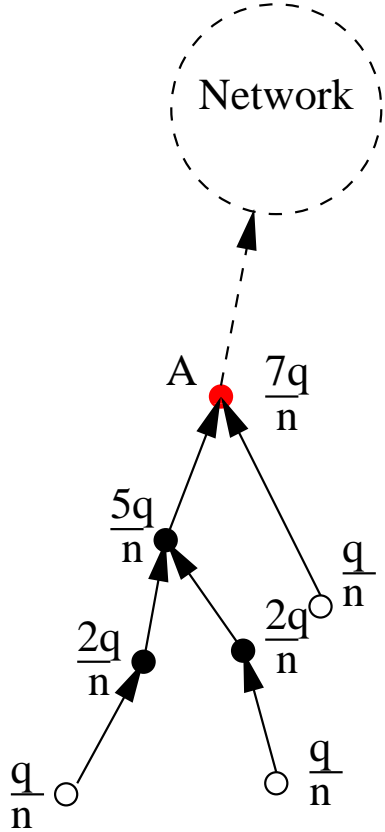}
\end{center}
    \caption{\label{fig2} Subgraph of a tree. A node A is shown 
together with all its predecessors.}
\end{figure}

\begin{figure}[t]
\begin{center}
    \includegraphics[width=7.5cm]{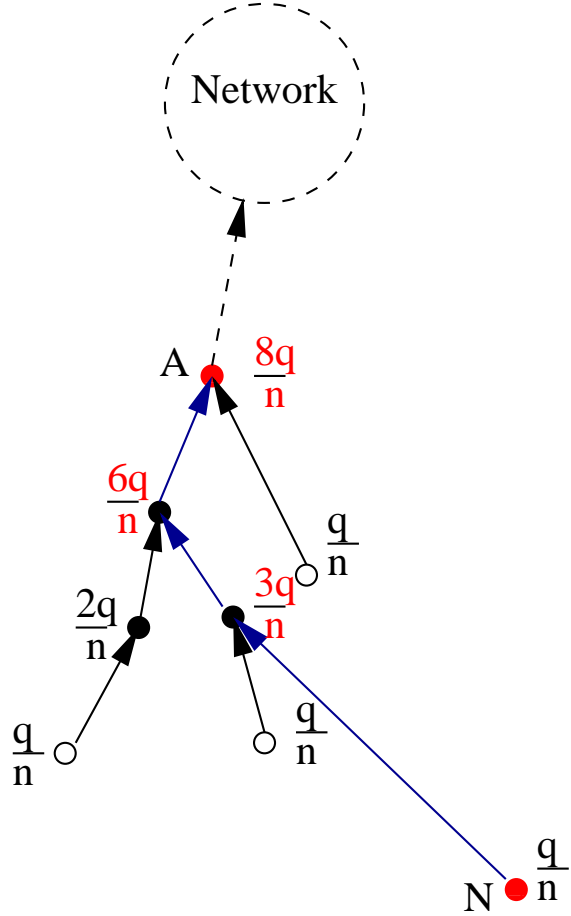}
\end{center}
    \caption{\label{fig3} If a new node N gives its link to any node of the subgraph, the 
PR of the uppermost node A will increase by $q/n$.}
\end{figure}

Let us suppose that $q$ is very small ($q\sim 0$) and can be treated as an infinitesimal. 
Eq.~(\ref{eq1}), to the first order in $q$, reads:
\begin{equation}
p(i)\sim\frac{q}{n}+\sum_{j: j \rightarrow i}\frac{p(j)}{k_{out}(j)}\hskip0.9cm i=1,2,
\dots, n
\label{eq3}
\end{equation}
where we have made the approximation $1-q\sim 1$. The general expression in Eq.~(\ref{equA})
grants that this approximation leads to the exact result.
Since $m=1$ there cannot be more than one path between two
nodes and the network is an oriented tree. 
Under the assumption that nodes have out-degree $1$, Eq.~(\ref{eq3}) reads:  
\begin{equation}
p(i)\sim\frac{q}{n}+\sum_{j: j \rightarrow i}p(j)\hskip0.9cm i=1,2,
\dots, n.
\label{eq4}
\end{equation}
meaning that the PR of a node is the sum of a constant term ($q/n$) and
the PR of its in-neighbors. In Fig.~(\ref{fig2})
we show a subgraph of a tree. Node $A$ is the root of the subgraph. A random walker 
moving from any node in the subtree and constrained by the directions of the links 
will necessarily reach $A$.
We call therefore the nodes in the subtree
\textit{predecessors} of $A$ (we include $A$ among its predecessors). 
The three empty circles are ``leaves'' of the subgraph, as 
they have no incoming links. Starting from the leaves, and using Eq.~(\ref{eq4})
recursively, it is possible to calculate the PR of all nodes of the diagram. The values are
reported next to the nodes. 
The figure shows that
\begin{itemize}
\item{all PR values
are multiples of the elementary unit $q/n$;}
\item{PR increases if one moves from a node to another 
by following a link;}
\item{the PR of each node $i$, in units of $q/n$, 
equals the number of its predecessors.}
\end{itemize}
In the following, PR is measured in units of $q/n$, and, accordingly, the probability distribution 
is written as $P_{PR}(l)$, with $l=1,2,...,n$. When a new node N gets connected to
a generic node of the subgraph of Fig.~(\ref{fig2}), the PR of 
node A increases by $q/n$. Further, all the nodes on the path between N and A count N as a predecessor
 and therefore they similarly increase their PR by $q/n$ (Fig.~(\ref{fig3})).  
In the next subsections we specialize the above to networks 
grown by a linear preferential attachment mechanism, either 
explicitly (Barab\'asi-Albert~\citep{BA} and
Dorogovtsev \textit{et al.}~\citep{DM}), or implicitly (Copying model~\citep{Copy}).

\subsection{Explicit preferential attachment}

In the model of Dorogovtsev \textit{et al.} (DMS)~\citep{DM}, adapted to a directed graph,
the probability that a new node $i$ 
attaches its link to a node $j$ (with in-degree $k_j$) is 
\begin{equation}
\Pi(k_j,a)=\frac{a+k_j}{\sum_{l=1}^{i-1}(a+k_l)}
\label{eq5}
\end{equation}
i.e. it only depends on the in-degree of the target node and on a real constant 
$a>0$. 
Eq.~(\ref{eq5}) is a generalization of the
linking probability of the Barab\'asi and Albert (BA) model~\citep{BA}, which is
recovered when $a$ coincides with the (fixed) out-degree $m$ of the nodes ($m=1$ in the present case).
The derivation below, therefore, encompasses  the BA model as a particular case. 

It is known that the DMS model leads to a scale-free in-degree distribution
with exponent $\gamma=2+a$. 
We start from a network with $n$ nodes. 
The probability distribution of PR is, initially, ${P_{PR}}^n(l)$.
In order to write a master equation that relates ${P_{PR}}^{n+1}(l)$ to 
${P_{PR}}^{n}(l)$, one notes that the addition of node $n+1$ increases by $q/n$ the 
PR of all nodes in the path between $n+1$ and $1$, while the others remain unaffected.
In this way, among the nodes of the path, PR $l-1$ will 
become $l$, whereas PR $l$ will become $l+1$.  
Let us consider a generic node $i$ with PR equal to $l$.
The probability $\Pi_i^n$ that the new link
will change the PR of $i$ from $l$ to $l+1$ is equal to the probability that
the link is received by any predecessor of $i$ (including $i$), i.e.
\begin{equation}
\Pi_i^n=\sum_{j=>i}\frac{a+k_j}{\sum_{t=1}^{n}(a+k_t)}
\label{eq6}
\end{equation}
where $j=>i$ indicates that $j$ is a predecessor of $i$.
Note that even if other predecessors of $i$ (besides $i$ itself) increase their
PR due to the attachment of the new node,
they cannot reach the value $l+1$, as their initial values are necessarily smaller than $l$.
Since all nodes have out-degree $m=1$, the total number of links
of a network with $n$ nodes is $n-1$ (we assume that the first node does not create links) 
and the denominator of 
Eq.~(\ref{eq6}) takes the simple form
\begin{equation}
\sum_{t=1}^{n}(a+k_t)=an+n-1=(a+1)n-1.
\label{eq7}
\end{equation}
The number of predecessors of $i$ is $l$, 
and the total number of adjacent links 
to the predecessors is $l-1$ (see Fig.~(\ref{fig2})).
One finally obtains:
\begin{equation}
\Pi_i^n=\sum_{j=>i}\frac{a+k_j}{(a+1)n-1}=\frac{(a+1)l-1}{(a+1)n-1}.
\label{eq8}
\end{equation}
The probability $\Pi^n(l)$ that the new link will alter the value of any node in 
the ``PR class'' $l$ is then:
\begin{equation}
\Pi^n(l)=nP_{PR}^{n}(l)\Pi_i^n=\frac{(a+1)l-1}{(a+1)-1/n}P_{PR}^{n}(l).
\label{eq9}
\end{equation}
The master equation then reads:
\begin{equation}
(n+1)P_{PR}^{n+1}(l)-nP_{PR}^{n}(l)=\Pi^n(l-1)-\Pi^n(l).
\label{eq10}
\end{equation}
Eq.~(\ref{eq10}) is a balance equation;
the left-hand side expresses the variation of the number of 
nodes in the ``PR class'' $l$, after the addition of the $(n+1)^{th}$ node.
The first term of the right-hand side is the probability that the introduction of the new node
increases the number of nodes in the ``PR class'' $l$ by one, the other term instead 
is the probability that a node leaves that class because its PR increases by one unit.
Since a single link is added at each iteration, only one node can make either transition,
so the right-hand side represents the expected variation in the population of nodes  
in the ``PR class'' $l$, i.e. exactly what we have on the left-hand side of Eq.~(\ref{eq10}).
 
\begin{figure}[t]
\begin{center}
    \includegraphics[width=9cm]{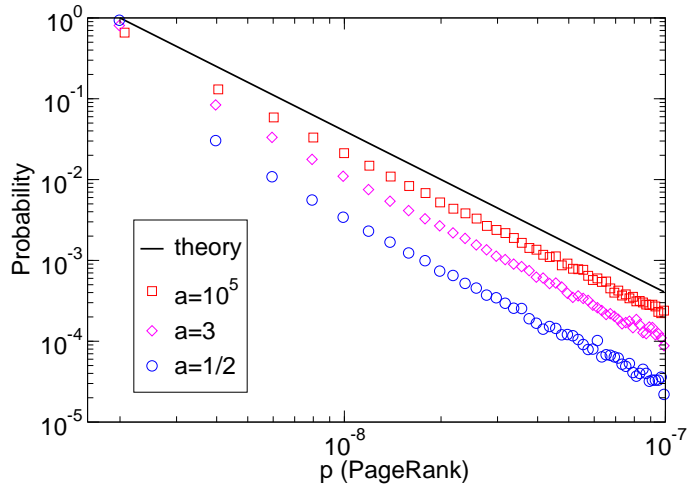}
\end{center}
    \caption{\label{fig41} Small-q PR distribution for DMS networks.}
\end{figure}

Note that Eq.~(\ref{eq10}) holds if $l>1$. When $l=1$, it must be modified, because
there are no nodes with zero PR and the first term on the right-hand-side would be ill-defined.
The modification, however, is simple. 
The new node $n+1$ is a ``leaf'', and it has PR $1$. At each iteration, therefore, 
the population of ``PR class'' $1$ is increased by one. We have
\begin{equation}
(n+1)P_{PR}^{n+1}(1)-nP_{PR}^{n}(1)=1-\Pi^n(1).
\label{eq11}
\end{equation}
We are interested in the stationary solutions of Eqs.~(\ref{eq10}) and (\ref{eq11}),
which can be derived by setting $P_{PR}^{n+1}(l)=P_{PR}^{n}(l)=P_{PR}(l)$ (valid in the limit
when $n\rightarrow\infty$). In this limit, one can safely neglect $1/n$ in Eq.~(\ref{eq9}).
After rearranging terms we obtain:
\begin{equation}
P_{PR}(l) = \left\{
  \begin{array}{ll}
\frac{(a+1)l-a-2}{(a+1)l+a}P_{PR}(l-1),
     & \mbox{\hskip1cm   if $l>1$;}\\
    \frac{a+1}{2a+1}, & \mbox{\hskip1cm   if $l=1$.}
  \end{array} \right.
\label{eq12}
\end{equation}
which leads to:
%\begin{equation}
%P_{PR}(l) = \frac{[(a+1)l-a-2][(a+1)l-2a-3][(a+1)l-3a-4]\cdot ...\cdot[(a+1)]}
%{[(a+1)l+a][(a+1)l-1][(a+1)l-a-2]\cdot ...\cdot[(2a+1)]}
%\label{eq13}
%\end{equation}
%
\begin{equation}
P_{PR}(l) = \frac{a(a+1)}{[(a+1)l+a][(a+1)l-1]}\sim \frac{1}{l^2}, \mbox{for $l\gg 1$}.
\label{eq13}
\end{equation}
The probability distribution of PR for a network built according to the
DMS model has a power law tail with exponent $\beta=2$, independently of $a$.
Fig.~(\ref{fig41}) shows PR distributions obtained from numerical simulations.  
They refer to three DMS networks, with parameter $a=1/2$, $3$, and $10^5$,
respectively. The number of nodes is $n=10^6$ and $q=0.001$. 
The tails of the three curves are straight lines in the double-logarithmic
scale of the plot, indicating a power law decay, and they are parallel.
The continuous line has the slope of the predicted trend, showing an excellent agreement.

As noted above, our analytical result and the simulation for $a=10^5$
shows that $\beta$ is independent of the
parameter $a$, surprisingly in contrast with what happens for the in-degree, 
that, in the limit $a\rightarrow\infty$, turns out to have an exponential distribution. 
The networks whose PR distributions are shown in the plot have been generated with $m=3$. 
Fig. (\ref{fig41}) then confirms that
our result holds even when $m>1$.

\subsection{Implicit preferential attachment: the Copying model}

The Copying model (CM)~\citep{Copy,redner} was originally introduced to
model the growth of the Web graph. It is based on the reasonable 
assumption that Web administrators, in creating a new page, often ``copy'' hyperlinks 
of pages they know. 
In this framework, a newly created node $i$ is a copy of a randomly chosen existing node 
$j$. This implies that $i$ sets links to all the neighbors of $j$.
Then, with probability $\alpha$, those links are rewired to other nodes, again chosen at random.
The model produces a scale-free network with a power law in-degree distribution characterized
by an exponent $\gamma=(2-\alpha)/(1-\alpha)$.

Although the linking mechanism is apparently unrelated to the degree of the target node,
a closer inspection
reveals that the copying mechanism implies an effective linear preferential 
attachment~\citep{Alebook}. 
To derive the PR distribution, we follow closely the strategy of the previous subsection. 

In order to affect the PR of a node $i$,
the link set by the new node $n+1$ must again attach to a predecessor of $i$.
It is useful to distinguish between the ``copying'' phase and the ``rewiring'' 
phase of the linking process.

In the copying phase, to affect the PR in $i$, the target node has to be a predecessor of $i$, 
excluding $i$ itself. After the rewiring phase, the node $i$ will avail itself of a new 
contribution in PR if the new link is untouched by the rewiring or rewired to another predecessor of 
$i$ (this time including $i$ itself).
Let's assume that node $i$ is originally in ``PR class'' $l$.
The probability to pick at random a predecessor of $i$ is $l/n$, if we include $i$, 
or $(l-1)/n$, if we exclude $i$. So, the probability $\Pi_i^n$ that the new 
link will change the PR of $i$ is:
\begin{equation}
\Pi_i^n=(1-\alpha)\,\frac{l-1}{n}+\alpha\,\frac{l}{n}=\frac{l+\alpha-1}{n}.
\label{eq14}
\end{equation}
The $\alpha$-dependent terms express the probability to have copying ($1-\alpha$) and
rewiring ($\alpha$). 
From Eq.~(\ref{eq14}) one can extend the result to all nodes with PR $l$, 
like in Eq.~(\ref{eq9})
\begin{equation}
\Pi^n(l)=nP_{PR}^{n}(l)\Pi_i^n=(l+\alpha-1)P_{PR}^{n}(l).
\label{eq15}
\end{equation}
Plugging the expression of $\Pi^n(l)$ in the 
balance equations (\ref{eq10}) and (\ref{eq11}), 
one obtains the following stationary solutions 
\begin{equation}
P_{PR}(l) = \left\{
  \begin{array}{ll}
\frac{l+\alpha-2}{l+\alpha}P_{PR}(l-1),
     & \mbox{\hskip1cm   if $l>1$;}\\
    \frac{1}{1+\alpha}, & \mbox{\hskip1cm   if $l=1$.}
  \end{array} \right.
\label{eq16}
\end{equation}
From the recursive relation of Eq.~(\ref{eq16}) the final expression for the
PR distribution follows
\begin{equation}
P_{PR}(l) = \frac{\alpha}{(l+\alpha)(l+\alpha-1)}\sim \frac{1}{l^2}, \mbox{for $l\gg 1$}.
\label{eq17}
\end{equation}
The result is analogous to the one obtained in the previous section. Since, as mentioned above, 
the linking mechanism of the CM hides an effective linear preferential attachment, the result
is not totally unexpected.

\begin{figure}[t]
\begin{center}
    \includegraphics[width=9cm]{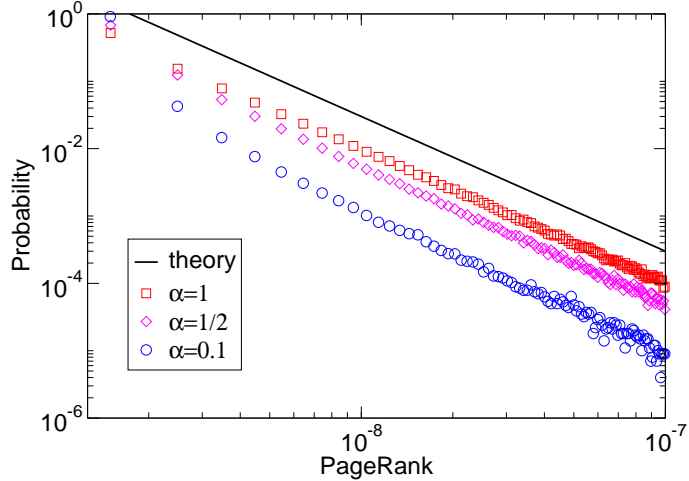}
\end{center}
    \caption{\label{fig42} Small-q PR distribution for CM networks.}
\end{figure}

A  numerical test of  the prediction in  Eq.~(\ref{eq17}) can be found in 
Fig.~(\ref{fig42}), where the PR distributions for three networks built with the CM, 
with $\alpha$ equal to $0.1$, $1/2$ and $1$, respectively, are shown. The other relevant parameters are 
$m=3$, $n=10^6$ and $q=0.001$. 
All the curves show the same slope (with exponent $2$) in a double-logarithmic plot.
Note that
the CM with $\alpha=1$ generates a network with an exponential in-degree distribution,
analogously to the DMS model in the limit $a\rightarrow\infty$. Again, this fact does not
affect the PR distribution.

\subsection{Hint to a general proof}
\label{hint}
We now hint to the possibility to extend the proof presented above to the case $m>1$. 
Let us work in the preferential attachment framework. Starting from Eq.~(\ref{equB}), we need to introduce 
the quantity $p_{ij}= \sum_{l^{ij} \in L^{ij} } ( \frac{ 1-q}{m} )^{d(l^{ij})} $, i.e. the contribution
of node $j$ to the PR in node $i$. This quantity, obviously, does not change in time. The addition of a new node at time $t$
(therefore the node is labelled $t$) contributes, on average, to the PR in $i$
\begin{equation}
p_{it}= \sum_{j \in n.n.i} p_{ij} \frac{ k_{j}(t) }{2mt}(1-q)
\end{equation}
where the sum runs over the nearest neighbors of the new node $t$ and $k_{j}(t)$ is the degree of node $j$ at time $t$.
Taking the average over all realizations of the process of growth and the limit for continuous time, 
one arrives to the following equation:
\begin{equation}
\label{equC}    
p(t_0,t)=\int_{t_0}^{t} p(t_0,s)k(s,t)\frac{(1-q)}{2mt}ds
\end{equation}
where $p(t_0,t)$ is the average contribution to the PR of a node born at time $t_0$ from
a node born at time $t$, and $k(s,t)$ is the average degree at time $t$ of a node born at time $s$.
If $p(t_0,t)$ can be explicitly found, then $p_T(t_0)$, the average PR at a generic time $T$ of a node born at 
time $t_0$ can be easily calculated as $p_T(t_0)=\int_{t_0}^T p(t_0,t) dt$. To compute $p(t_0,t)$ we need $k(s,t)$ first.
In the context of preferential attachment $k(s,t)$ is found to be $k(s,t)=m(t/t_0)^{1/2}$ (this easily follows from
$dk/dt  = k/2mt$ and $k(t_0,t_0)=m$). This expression provides
the kernel for the integral equation~(\ref{equC}). Once Eq.~(\ref{equC}) is solved
(taking into account the correct boundary condition $p(t_0,t_0)=1$)
and the result properly integrated,
it gives, in the limit $T>>t_0$ and $q \rightarrow 0$, $p_T(t_0) \propto T^{3/2}/t_0$, which in turn gives
the expected result: PR is algebraically distributed with an exponent equal to $2$ independently of $m$.
In general the kernel $k(s,t)/2tm$ in~(\ref{equC}) needs to be replaced by that appropriate
to the growth model under consideration.

\subsection{Beyond preferential attachment}

We have seen that the PR distribution for special networks has a power law
tail with exponent $2$, independently of the in-degree distribution of the network, which
needs not even be a power law (e.g. DMS model for $a\rightarrow\infty$, 
CM for $\alpha\rightarrow 1$). This evidence, together with the observation
that the PR distribution for the real Web (where a relatively small $q$ is usually employed)
has also a power law distribution with exponent close to $2$, may erroneously
lead to the conclusion that the above result applies to a general graph.  

%One might then wonder whether this result holds for an arbitrary graph. 
%We believe that the tree-like structure we have so far considered,
%where any simple random walk ends up at the root node(s), 
%is essential for the result to hold. In fact, if there is a cascade towards
%a root, the distribution of PR will be broad, as the PR values of older nodes keep
%increasing, due to the addition of new nodes, whereas the newest among the nodes 
%will all carry the entry value $q/n$. 

\begin{figure}[t]
\begin{center}
    \includegraphics[width=9cm]{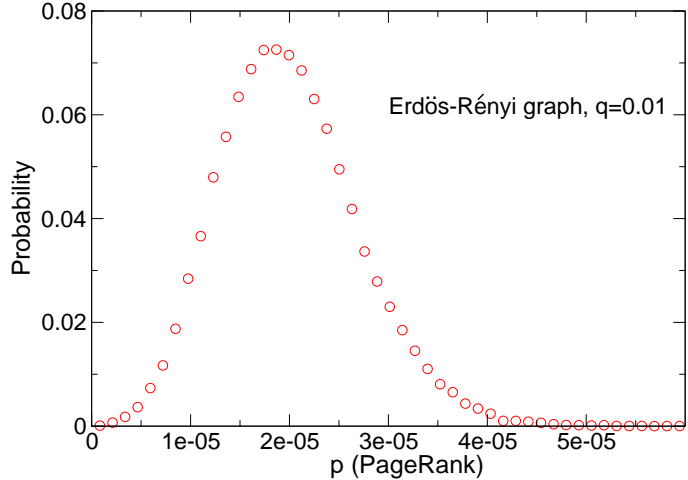}
\end{center}
    \caption{\label{fig4} Small-q PR distribution for an Erd\"os-R\'enyi random graph.}
\end{figure}

A  numerical test on a random graph \textit{a la} Erd\"os-R\'enyi~\citep{erdos} 
shows the limits of the validity of our result.
An Erd\"os-R\'enyi graph is built
starting from a set of $n$ nodes, and setting a link independently and with a probability $r$ between
any pair of nodes. The resulting network has a Poissonian 
degree distribution, with mean $rn$. In order to make the graph directed,
we orient the link $i-j$ with equal probability  
from $i$ to $j$ or from $j$ to $i$. There is no ``center'' and no PR flux
towards a core of nodes, unlike the networks we have studied above. All nodes will thus 
have equal rights, and we expect little differences in their PR values. 
Fig.~(\ref{fig4}) shows the 
PR distribution for a random graph with $50000$ nodes and $r=0.0002$; the damping
factor $q$ is $0.01$. The distribution appears to be a Poissonian, like that of in-degree. 

\begin{figure}[t]
\begin{center}
    \includegraphics[width=9cm]{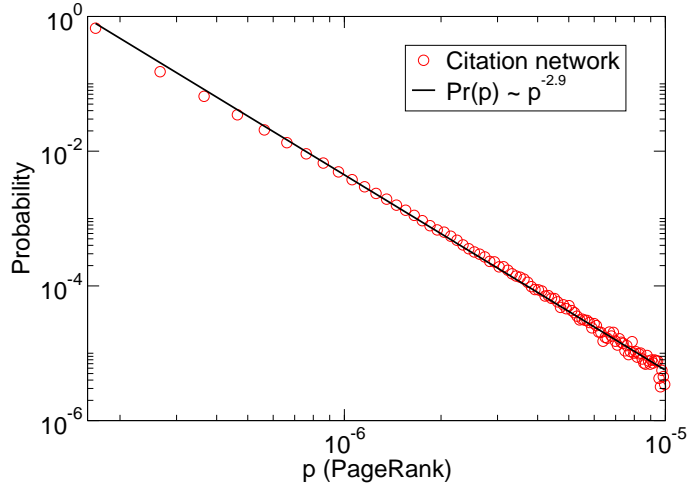}
\end{center}
    \caption{\label{fig4cit} Small-q PR distribution for a citation network of U.~S. patents.
The continuous line is a power law fit of the tail.}
\end{figure}

It would be interesting 
to understand whether the result presented in this paper 
holds for all networks in which random walkers stream 
towards a core of nodes. We expect the PR distribution 
to be a power law quite generally, but we have no arguments
hinting to a universal occurrence of the exponent $2$. 
Numerical evidences suggest, in fact, that other exponents are possible. In Fig.~(\ref{fig4cit}) 
we show the small-q PR distribution for a citation network of U.S. patents ($q=0.001$).
Citation networks are practical examples of the directed trees we have analyzed
so far, as a new paper must necessarily cite older papers.  
The data~\citep{patent} refer to over 3 million U.S. 
patents granted between January 1963 and December 1999, and comprise
all citations made to these patents between 1975 and 1999. 
The PR distribution is skewed, as expected, but 
the slope of the tail is quite different from $2$, being close to $3$.

\section{The limit $q\rightarrow 1$}
\label{seclim1}

When $q=1$ all nodes have the same PR value $1/n$.
In the following we study the limit  $q\rightarrow1$ but $q\neq 1$.
In our Eq.~(\ref{eq1}), the constant $q/n\sim 1/n$ is now much larger than 
the sum on the right-hand-side (we treat $1-q$ as an
infinitesimal). The PR distribution will then be very narrow 
and squeezed towards $q/n$, which is not interesting.
However, the sum over the neighbors in Eq.~(\ref{eq1})
determines the variable contribution to PR,
which is responsible for the differences in PR between the nodes.
Therefore, we isolate this piece, and call it \textit{reduced PageRank} (RPR). So, 
the RPR $p_r(i)$ of a node $i$ is defined as
\begin{equation}
p_r(i)=p(i)-\frac{q}{n}\hskip0.9cm i=1,2,
\dots, n.
\label{eq23}
\end{equation}
The RPR is the probability that, during the PR process, 
a node is visited by a walker coming through 
any of its incoming links.
One can show that the distribution of RPR coincides with the in-degree distribution
on every graph, provided the out-degree is a constant $m$.
In this case, in fact,  
when we replace PR with RPR through the relation (\ref{eq23}),
Eq.~(\ref{eq1}) assumes the following form
\begin{eqnarray}
p_r(i)&=&\frac{1-q}{m}\sum_{j: j \rightarrow i}[p_r(j)+q/n] \nonumber\\
&=&\frac{q(1-q)}{mn}k_{in}(i)+\frac{1-q}{m}\sum_{j: j \rightarrow i}p_r(j),\\
\nonumber
\label{eq24}
\end{eqnarray}
where $k_{in}(i)$ is the in-degree of $i$. From Eq.~(19) it follows that
the RPR of a node is of order $1-q$. All terms coming from the 
sum are of order $(1-q)^2$ and can be safely neglected. Finally,
\begin{equation}
p_r(i)\sim\frac{q(1-q)}{mn}k_{in}(i),\hskip0.9cm i=1,2,
\dots, n. 
\label{eq25}
\end{equation}
The RPR of a node is then proportional to its in-degree, and  the corresponding  
distributions coincide, under no assumptions other than the out-degree
is a constant. Therefore, the  result has a wide generality. It is also intuitive 
how to extend it to the case in which the out-degree is not constant but approximately
the same for all nodes. Out-degree distributions concentrated about some value, like Gaussians,
Poissonians, exponentials, etc., should not change the result.

\begin{figure}[t]
\begin{center}
    \includegraphics[width=9cm]{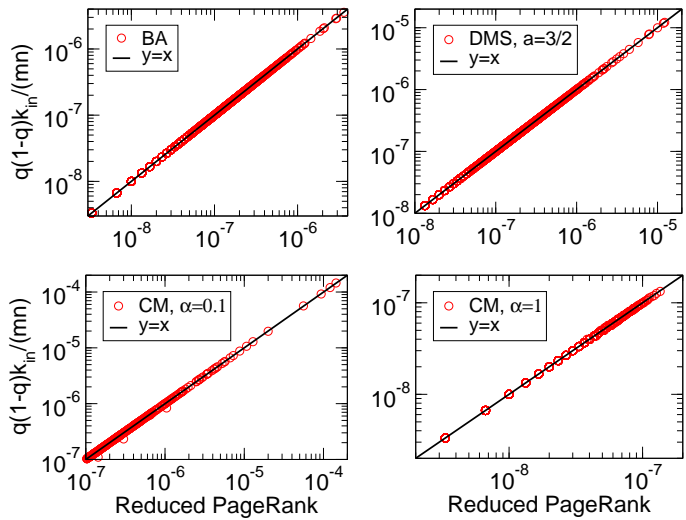}
\end{center}
    \caption{\label{fig48} Numerical test of Eq.~(\ref{eq25}) for networks
built according to different growth models. The number of nodes is $n=10^5$, $m=3$ and $q=0.999$. 
Top left: BA model. Top right: DMS model for $a=3/2$.
Bottom left: CM for $\alpha=0.1$. Bottom right: CM for $\alpha=1$.} 
 \end{figure}

Fig.~(\ref{fig48}) shows a test of Eq.~(\ref{eq25}). Each of the four plots 
is a scatter plot relative to a different network;
three of them are scale-free and one has an exponential in-degree distribution (bottom right), as it 
has been generated with a CM process for $\alpha=1$.
The RPR of a generic node is compared with 
the right-hand-side of Eq.~(\ref{eq25}). The continuous line 
represents the equality of the two variables. The comparison with the 
data points is excellent in all cases. 

After the submission of the paper we realized that the result
of this section had already been derived and presented in \citep{gems}. We apologize 
with Chen and colleagues for the unfortunate accident.

\section{Conclusions}

Since the birth of Google, PR has attracted a lot of interest 
from the scientific community, but the deep reasons behind its capacity 
to capture the ``quality'' better than other and more used topological descriptors (e.g. in-degree)
are not yet clear. 
We studied PR in a more general framework than its original field of application (the Web graph). 
We derived some exact results for PR distributions in the 
limit when the damping factor $q$ approaches the two extreme values $0$ and $1$.
When $q\rightarrow 0$, for networks 
without directed loops and where 
walkers stream towards a central core of nodes (roots),
PR can be in principle calculated in a single sweep over the nodes, starting from 
the leaves and converging shell-wise towards the center. 
This feature 
allowed us to calculate exactly the distribution of PR for networks
built according to some peculiar linking strategies, like that of the DMS model 
(which includes the BA model as a special case) and of the CM.
In these cases, the PR distribution has a power law tail with exponent $2$,
for any choice of the model parameters, that, on the contrary, strongly affect the
in-degree distribution. This possibly suggests that the PR process allows to 
diversify the roles of the different nodes much more than in-degree, and it is a better 
criterion to rank nodes.
Many networks have the features that grant, on a first approximation, 
the applicability of our results.
Networks grown about one or more centers, with new nodes pointing
mostly to older nodes belong to this class. The Web itself could be taken as an example 
of this kind of networks.
The PR distribution of the Web graph is usually calculated for $q=0.15$, which is quite close to zero,
showing an exponent indeed very close to $2$ (see Fig.~(\ref{figPR})). 
Work is in progress to determine what are the broadest conditions that yield
this ``universal'' behavior.

In the limit $q\rightarrow 1$, PR is a linear function of in-degree, as long 
as the out-degree of the nodes is fixed. The relation holds at the level of the single node,
and not merely in the statistical sense. We plan to investigate how general this result is by
relaxing the assumption of constant out-degree and trying various distributions. 

To summarize, the PR distribution strongly depends on the 
value of the damping factor $q$, is in general ``uncorrelated'' from 
the corresponding in-degree distribution, but depends on the overall topological
organizaton of the graph. This is
not in contradiction with the findings of~\citep{santoMF}, where 
a correlation between the two variables was observed, because the correlation 
involves the in-degree and the mean PR-value of all nodes with that in-degree. 
Within each in-degree class PR has large fluctuations.

\section*{Acknowledgments}

S.F. acknowledges financial support from 
the Volkswagen Foundation and the NSF IIS-0513650 award.

%%%%%%%%%%%%%%%%%%%%%%%%%%%%%%%%%%%%%%%%%%%%%%%%%%%%%%%%%%%%%%

\end{document}